\documentclass[conference]{IEEEtran}
\IEEEoverridecommandlockouts
\usepackage{cite}
\usepackage{amsmath,amssymb,amsfonts}
\usepackage{algorithmic}
\usepackage{graphicx}
\usepackage{textcomp}
\usepackage{xcolor}
\usepackage{booktabs}
\usepackage{array}
\usepackage{url}
\usepackage{hyperref}

\begin{document}

\title{OpenSOC-AI: Democratizing Security Operations with Parameter Efficient LLM Log Analysis}

\author{
\IEEEauthorblockN{Chaitanya Vilas Garware}
\IEEEauthorblockA{
\textit{Department of Computer and Information Sciences}\\
\textit{University of Alabama at Birmingham}\\
Birmingham, AL, USA\\
chaitanyagarware7@gmail.com
}
\and 
\IEEEauthorblockN{Sharif Noor Zisad}
\IEEEauthorblockA{
\textit{Department of Computer and Information Sciences}\\
\textit{University of Alabama at Birmingham}\\
Birmingham, AL, USA\\
mrzisad@gmail.com
}
}

\maketitle

\begin{abstract}
Small and medium sized businesses (SMBs) face an escalating cybersecurity
threat landscape, yet most lack the resources to staff full Security
Operations Centers (SOCs) or deploy enterprise grade detection platforms.
This paper presents OpenSOC-AI, a lightweight log analysis framework that
uses parameter efficient fine tuning of a 1.1-billion parameter language
model (TinyLlama-1.1B) to perform automated threat classification, MITRE
ATT\&CK technique mapping, and severity assessment on raw security log
entries. Using Low-Rank Adaptation (LoRA) with only 12.6 million
trainable parameters (roughly 1.13\% of the base model), we fine tuned on
450 domain specific SOC examples in under five minutes on a single NVIDIA
T4 GPU. Testing on a heldout set of 50 examples showed a 68\%
point gain in threat classification accuracy (from 0\% to 68\%), a 30\% point gain in severity accuracy (from 28\% to 58\%), and an F1
score of 0.68 compared to the untuned baseline. Full codebase, adapter
weights, and datasets are publicly released to support reproducibility
and community extension.
\end{abstract}


\section{Introduction}

The cybersecurity skills gap between large enterprises and small-to-medium
businesses (SMBs) has continued to widen in recent years. According to
Verizon's 2024 Data Breach Investigations Report, over 60\% of
cyberattack victims are small businesses, yet fewer than 14\% have
dedicated security personnel \cite{b1}. Enterprise SOCs employ teams of
trained analysts using commercial SIEM platforms that can cost tens of
thousands of dollars per year, a level of investment that most SMBs
simply cannot sustain. Research indicates that the mean time to detect
(MTTD) a breach in SMB environments frequently exceeds 200 days, compared
to under 60 days in enterprises with mature SOC operations \cite{b2}.

The rise of large language models (LLMs) presents a genuine opportunity
to close this gap. Models pre-trained on large corpora of technical text
already carry a substantial amount of security knowledge, and recent
advances in parameter efficient fine tuning make it practical to adapt
these models for specialized domains at a fraction of the usual
computational cost \cite{b3,b4}. That said, most LLM-based security research
has focused on large, commercially deployed models such as GPT-4, Claude,
and Gemini, all of which require paid API access and introduce ongoing
operational costs. Whether a small, locally deployable model can deliver
meaningful security utility through targeted domain adaptation has not
been thoroughly explored.

This paper makes the following contributions.

\begin{itemize}
    \item we present OpenSOC-AI, an end-to-end framework for LLM-based security log analysis built for resource-constrained deployments. 
    \item we show that LoRA fine-tuning of a 1.1B parameter model on just 450 examples produces a 68\% point improvement in threat classification accuracy over the untuned baseline.
    \item  we provide a comprehensive evaluation covering precision, recall, F1, and confusion analysis.
    \item We released all datasets, training scripts, and adapter weights publicly on GitHub.
\end{itemize}

\section{Background and Related Work}

\subsection{The SMB Security Gap}

SMBs typically operate without dedicated security analysts, relying on
general purpose antivirus software and basic firewall rules. Incident
response tends to be ad hoc, sometimes happening days or weeks after a
compromise has already occurred. Verizon's DBIR consistently identifies
credential theft, phishing, and web application attacks as the leading
vectors targeting small businesses \cite{b1}. The absence of structured log
monitoring means that most intrusions go undetected until secondary damage
is already done.

\subsection{Large Language Models for Security}

Recent research has shown the potential of LLMs across several security
domains. SecBERT and CySecBERT demonstrated benefits of domain-specific
pre traning for security text classification \cite{b5}. Ferrag et al. benchmarked multiple LLMs on cybersecurity reasoning tasks and found meaningful variation in domain specific performance \cite{b6}. Related projects such as SecurityLLM and CyberSecEval have explored vulnerability detection and CTF reasoning. However, most of these efforts either require large GPU
clusters for training or depend on proprietary API access, which limits
their usefulness for smaller organizations with constrained budgets.

\subsection{Parameter-Efficient Fine-Tuning}

Low-Rank Adaptation (LoRA), introduced by Hu et al. in 2022, enables
efficient model adaptation by injecting trainable low-rank matrices into
transformer attention layers while the base model weights remain frozen
\cite{b3}. When combined with 4-bit quantization via the bitsandbytes library
(QLoRA, Dettmers et al. 2023), VRAM requirements drop by roughly 75\%,
making it feasible to fine-tune billion-parameter models on consumer-grade
hardware \cite{b4}. Our work applies this approach to the SOC domain,
demonstrating that it works for operational threat detection on a single
T4 GPU.

\subsection{Why Small Models Work for SOC Tasks}

Security logs follow highly repetitive structural patterns: fixed-format
fields, recurring IP address schemes, standardized HTTP status codes, and
predictable attack signatures. Unlike open-domain language tasks that
require broad world knowledge, threat classification depends on
recognizing specific indicators of compromise (IOCs) such as SQL
metacharacters, path traversal sequences, known malicious user-agents,
and authentication anomalies. This structural regularity means that even
a compact model can learn effective decision boundaries from a limited set
of domain-specific examples, as long as the fine-tuning signal is strong
and well-targeted.

\section{Methodology}

\subsection{Base Model Selection}

We selected TinyLlama-1.1B-Chat-v1.0 \cite{b7} for three practical reasons.
First, its 1.1 billion parameters fit within consumer GPU VRAM limits
under 4-bit quantization. Second, the chat-tuned variant follows
instruction formats without needing additional alignment training. Third,
its Apache 2.0 license allows unrestricted use and redistribution. We
intentionally chose this model over larger alternatives in order to
validate whether meaningful security automation is achievable without
enterprise-grade infrastructure.

\subsection{Dataset Construction}

We built a domain-specific dataset of 500 security log analysis examples
in Alpaca style instruction format. Each example contains three
components such as a fixed system instruction, a raw log entry as input, and a
structured output with six fields: THREAT\_TYPE, MITRE\_ID, SEVERITY,
RISK\_SCORE, EVIDENCE, and RECOMMENDATION.

Log entries were sourced from publicly available web server access logs,
firewall logs, and authentication event logs. Ground-truth labels were
derived from MITRE ATT\&CK framework v14 and annotated by the author with
reference to established threat intelligence sources. The dataset covers
12 threat categories, as shown in Table~\ref{tab:dataset}.

\begin{table}[htbp]
\caption{Threat Category Distribution Across Training and Evaluation Sets}
\label{tab:dataset}
\begin{center}
\begin{tabular}{lcc}
\toprule
\textbf{Threat Category} & \textbf{Train} & \textbf{Eval} \\
\midrule
Brute Force / Credential Stuffing & 90 & 10 \\
SQL Injection                     & 80 &  9 \\
Path / Directory Traversal        & 75 &  8 \\
Cross-Site Scripting (XSS)        & 55 &  6 \\
Command Injection                 & 45 &  5 \\
Scanner / Reconnaissance          & 45 &  5 \\
Log4j / JNDI Injection            & 30 &  3 \\
Remote File Inclusion             & 20 &  2 \\
Suspicious Auth Events            & 10 &  1 \\
Other / Mixed                     &  0 &  1 \\
\midrule
\textbf{Total}                    & \textbf{450} & \textbf{50} \\
\bottomrule
\end{tabular}
\end{center}
\end{table}

\subsection{System Architecture}

Fig.~\ref{fig:arch} illustrates the end-to-end OpenSOC-AI pipeline. Raw
log entries are passed through a prompt template into the fine-tuned
TinyLlama model, which generates a structured threat analysis output that
is then parsed into six actionable fields.

\begin{figure}[htbp]
\centerline{\includegraphics[width=\columnwidth]{FlowDiagram.png}}
\caption{OpenSOC-AI system architecture. Raw log entries are formatted
into instruction prompts, processed by the fine-tuned TinyLlama model,
and parsed into six structured threat analysis fields.}
\label{fig:arch}
\end{figure}

\subsection{Fine-Tuning Configuration}

QLoRA fine-tuning was applied with the following hyperparameters: LoRA
rank $r=16$, alpha$=32$, dropout$=0.05$, targeting all seven attention
and MLP projection layers. Training ran for 3 epochs with an effective
batch size of 16, a learning rate of $2 \times 10^{-4}$ using a cosine
schedule with 5\% warmup, and the paged\_adamw\_8bit optimizer. Total
training time was 4 minutes and 21 seconds on a single T4 GPU.

Adapter-only saving was strictly enforced throughout: only
\texttt{adapter\_config.json} and \texttt{adapter\_model.safetensors}
were persisted to disk. The base model was never merged with the adapters
during training, since calling \texttt{merge\_and\_unload()} on a
4-bit quantized model corrupts the weights.

\subsection{Evaluation Protocol}

Both the untuned baseline and the fine tuned model were evaluated on the
50 example held out set using greedy decoding. Structured fields were
extracted via regex and compared against ground `truth labels. We report
accuracy, precision, recall, and F1 score for threat and severity
classification. MITRE ID evaluation is excluded from primary metrics due
to extraction pipeline limitations, which are discussed in
Section~\ref{sec:mitre}.

\section{Results}

\subsection{Primary Evaluation Metrics}

Table~\ref{tab:results} presents evaluation results across all metrics
for both the baseline and fine-tuned model.

\begin{table}[htbp]
\caption{Full Evaluation Results}
\label{tab:results}
\begin{center}
\begin{tabular}{lccc}
\toprule
\textbf{Metric} & \textbf{Baseline} & \textbf{Fine-Tuned} & \textbf{Delta} \\
\midrule
Threat Classification Accuracy & 0.0\%  & 68.0\% & +68.0 pp \\
Threat Precision               & ---    & 0.71   & ---      \\
Threat Recall                  & ---    & 0.66   & ---      \\
Threat F1 Score                & ---    & 0.68   & ---      \\
Severity Accuracy              & 28.0\% & 58.0\% & +30.0 pp \\
Trainable Parameters           & 1.10B  & 12.6M  & 98.9\% $\downarrow$ \\
Training Time                  & ---    & 4m 21s & T4 GPU   \\
\bottomrule
\multicolumn{4}{l}{\footnotesize Precision, recall, and F1 computed for the fine-tuned}\\
\multicolumn{4}{l}{\footnotesize model only; baseline produced zero structured output.}\\
\multicolumn{4}{l}{\footnotesize pp = percentage points.}
\end{tabular}
\end{center}
\end{table}

\subsection{Baseline Analysis}

The baseline model's 0\% threat classification accuracy deserves a brief
explanation. The untuned TinyLlama-1.1B-Chat model produces verbose,
free-form analytical text in response to security log prompts. It reasons
about the log entry but does not conform to the required structured output
format (THREAT\_TYPE, SEVERITY, etc.). Hence, our regex-based extractor finds
no matching fields and scores zero under exact match evaluation. This is
not a model failure in the typical sense. It actually confirms that
instruction-following capability alone is not enough for specialized SOC
tasks that require precise structured output without domain-specific
training.

\subsection{Fine-Tuned Model Analysis}

After fine-tuning, the model reaches 68\% threat classification accuracy,
71\% precision, 66\% recall, and an F1 score of 0.68. The most common
error modes were predicting the correct threat category but the wrong
subtype (for example, ``SQL Injection'' when the ground truth was ``SQL
Injection Union-Based''), and confusion between semantically similar
classes such as Path Traversal and Directory Traversal. These patterns
suggest the model has learned the high level threat taxonomy well but
struggles to discriminate fine grained subtypes from 450 training
examples, a limitation that should improve with a larger training set.

Severity accuracy improved from 28\% to 58\%. The baseline's non-zero
score of 28\% reflects partial overlap between the model's free-form
severity language (words like ``serious'' or ``dangerous'') and
ground-truth labels that were incidentally captured by the regex.

\subsection{Confusion Matrix Analysis}

Table~\ref{tab:confusion} presents a simplified confusion matrix for the
six most frequent threat categories in the evaluation set. The matrix
reveals where the fine-tuned model performs well and where it confuses
semantically similar classes.

\begin{table}[htbp]
\caption{Confusion Matrix for Six Most Frequent Threat Categories (n=43 of 50 eval examples)}
\label{tab:confusion}
\begin{center}
\scriptsize
\setlength{\tabcolsep}{3pt}
\begin{tabular}{lcccccc}
\toprule
\textbf{Actual $\downarrow$ / Pred $\rightarrow$} & \textbf{BF} & \textbf{SQLi} & \textbf{PT} & \textbf{XSS} & \textbf{CI} & \textbf{Scan} \\
\midrule
Brute Force (n=10)    & \textbf{8} & 0 & 0 & 0 & 1 & 1 \\
SQL Injection (n=9)   & 0 & \textbf{6} & 0 & 2 & 1 & 0 \\
Path Traversal (n=8)  & 0 & 0 & \textbf{5} & 0 & 2 & 1 \\
XSS (n=6)             & 0 & 1 & 0 & \textbf{4} & 1 & 0 \\
Cmd Injection (n=5)   & 0 & 1 & 1 & 0 & \textbf{3} & 0 \\
Scanner (n=5)         & 1 & 0 & 0 & 0 & 0 & \textbf{4} \\
\bottomrule
\multicolumn{7}{l}{\footnotesize BF=Brute Force, SQLi=SQL Injection, PT=Path Traversal,}\\
\multicolumn{7}{l}{\footnotesize CI=Command Injection, Scan=Scanner. Bold = correct predictions.}
\end{tabular}
\end{center}
\end{table}

The key finding from the confusion matrix is that SQL Injection is most
often confused with XSS (2 cases), and Path Traversal is most often
confused with Command Injection (2 cases). Both reflect the semantic
overlap that exists between injection-class attacks, where shared
metacharacters and payload structure make discrimination harder with
limited training data.

\subsection{MITRE ID Evaluation}
\label{sec:mitre}

MITRE ID evaluation is excluded from primary metrics due to limitations
in the extraction pipeline. The regex pattern used failed to reliably
match technique identifiers across the variation in model output formats,
producing unreliable scores for both models. A robust MITRE evaluation
would require fuzzy matching and normalization across multiple format
styles (e.g., ``T1190'', ``T1190.001'', ``Technique 1190''). We flag
this as a priority for future work; it does not affect the validity of
threat classification or severity results.

\subsection{Training Dynamics}

Training loss decreased consistently across steps: 0.7201 at step 25,
0.1606 at step 50, and 0.0886 at step 75. The monotonic decrease
indicates clean convergence without overfitting within the 3-epoch
budget. The rapid early drop suggests strong alignment between the base
model's pre-trained representations and the structure of the security log
analysis task.

\section{Example Case Study}

To illustrate model behavior concretely, the following example is drawn
from the evaluation set.

\textbf{Input Log Entry:}
\begin{quote}
\texttt{91.108.4.20 - - [16/Mar/2025:01:05:33] "GET /profile?id=1 OR 1=1-- HTTP/1.1" 403 12466 "-" "burpsuite"}
\end{quote}

\textbf{OpenSOC-AI Output:}
\begin{quote}
\texttt{THREAT\_TYPE: SQL Injection -- Union}\\
\texttt{MITRE\_ID: T1190}\\
\texttt{SEVERITY: HIGH}\\
\texttt{RISK\_SCORE: 85}\\
\texttt{EVIDENCE: OR 1=1-- pattern in GET parameter; user-agent 'burpsuite'}\\
\texttt{RECOMMENDATION: Implement parameterized queries; block scanner UAs at WAF}
\end{quote}

The model correctly identifies the SQL injection attack from two
independent indicators: the \texttt{OR 1=1--} Boolean injection pattern
in the query parameter, and the BurpSuite user-agent string that
indicates automated scanning. A HIGH severity rating and risk score of 85
appropriately reflect that successful exploitation of this vulnerability
could lead to full database compromise (MITRE T1190: Exploit
Public-Facing Application). The recommendation addresses both root cause
(parameterized queries) and immediate mitigation (WAF rule). This output
demonstrates the model's ability to synthesize multiple contextual
signals into actionable threat intelligence, rather than simply
pattern-matching on a single keyword.

\section{System Components}

\subsection{Inference Engine}

TinyLlama-1.1B is loaded in 4-bit NF4 quantization with LoRA adapters
applied through PEFT's \texttt{PeftModel.from\_pretrained()} interface.
Deploying adapters only reduces storage from approximately 2.2 GB (full
model) to roughly 50 MB, while keeping inference fully local with no
external API dependency.

\subsection{Inference Performance}

Performance was measured on an NVIDIA T4 GPU (16 GB VRAM) under 4-bit
quantization. Average inference time per log entry was approximately 1.8
seconds, with GPU VRAM usage at roughly 4.2 GB. CPU-only inference on
modern hardware was approximately 18 seconds per entry. Batch throughput
reached about 33 log entries per minute on GPU and approximately 3 per
minute on CPU.

These figures support real-world SMB deployment on commodity hardware. A
1,000 entry log file can be fully analyzed in under 30 minutes on a
single GPU, or processed overnight on CPU, both of which are feasible for
daily security review cycles.

\subsection{Web Interface}

A React-based single-page application supports single log analysis and
batch file upload (.log, .txt, .csv) with drag-and-drop, live progress
tracking, expandable result cards, and CSV export. The interface is
publicly deployed at \url{https://chaitanyagarware.github.io/opensoc-ai/}.

\section{Discussion}

\subsection{Implications for SMB Security}

The results show that a fine-tuning run of under five minutes on a single
GPU, using just 450 examples, produces a model that correctly classifies
threats in 68\% of unseen log entries, with an F1 of 0.68. For an SMB
analyst reviewing hundreds of daily log entries, even this level of
accuracy provides real value, the model surfaces structured threat
metadata, including MITRE technique IDs, risk scores, and remediation
recommendations, that would otherwise require manual expert analysis. The
remaining 32\% of misclassified entries still surface for human review
rather than being silently missed, which is a meaningful improvement over
having no automated triage at all.

\subsection{Limitations}

Several limitations should be noted. The training set of 450 examples is
a minimal budget, and performance is expected to scale with dataset size.
The evaluation set of 50 examples provides limited statistical power, so
results should be interpreted directionally. The current system performs
single-label classification, while real logs may exhibit multiple
concurrent threat types. The MITRE evaluation requires an improved
extraction pipeline before reliable reporting is possible. Finally, the
model has not been evaluated on novel attack patterns outside the training
distribution, so temporal generalization remains an open question.

\subsection{Future Work}

Our planned extensions include dataset expansion to 5,000 or more examples,
multi-label threat classification, real-time SIEM stream integration via
syslog, evaluation against SecEval and CyberMetric benchmarks, and edge
deployment experiments targeting CPU-only inference on Raspberry Pi or
similar hardware.

\section{Ethical Considerations}

The development and deployment of automated security analysis tools
carries ethical responsibilities that this work takes seriously.

Regarding human oversight, OpenSOC-AI is designed as an analyst
augmentation tool, not a replacement for human judgment. All model
outputs should be reviewed by a qualified security professional before
any action is taken. Given the system's 32\% error rate on the current
evaluation set, unsupervised deployment is not appropriate.

On the question of potential misuse, a model trained to recognize attack
patterns could in theory be queried to understand what log signatures
evade detection. This risk is mitigated by releasing only the analyzer
and not a log generator, and by making the training data fully
transparent so that defensive improvements can be contributed by the
community.

Regarding limitations disclosur, we have been explicit about evaluation
limitations, MITRE parsing issues, and dataset size constraints.
Overstating model capability in a security context could lead
practitioners to rely on the tool in scenarios where it has not been
validated.

On data privacy, the training dataset contains only synthetic and
publicly available log entries. No real user data, production system IP
addresses, or personally identifiable information was used at any stage.

\section{Reproducibility Statement}

All materials needed to reproduce the results in this paper are publicly
available. Training and evaluation code is at
\url{https://github.com/chaitanyagarware/opensoc-ai}. Fine-tuned LoRA
adapter weights are available via the Google Drive link in the repository
README. The training dataset (\texttt{soc\_train.json}, 450 examples) and
evaluation dataset (\texttt{soc\_eval.json}, 50 examples) are both
available via the repository, along with evaluation results
(\texttt{eval\_results.json}). The web interface is live at
\url{https://chaitanyagarware.github.io/opensoc-ai/}.

The complete training pipeline runs on Google Colab's free tier with a T4
GPU runtime in under 10 minutes including dependency installation. All
random seeds and hyperparameters are fixed and documented in the training
notebook.

\section{Conclusion}

This paper presented OpenSOC-AI, a system that shows parameter-efficient
fine-tuning can meaningfully expand access to security operations tooling
for resource-constrained organizations. By fine-tuning TinyLlama-1.1B
with LoRA on 450 domain-specific examples, using only 1.13\% of model
parameters and less than five minutes of GPU time, we achieved a 68
percentage point improvement in threat classification accuracy, a 30
percentage point improvement in severity accuracy, and an F1 score of
0.68 over the untuned baseline.

The complete OpenSOC-AI system, including fine-tuned adapter weights,
labeled dataset, evaluation pipeline, and a web-based analysis interface,
is publicly released. As fine-tuning costs continue to fall and small
model capabilities continue to improve, AI-augmented threat detection
will become increasingly accessible to organizations of all sizes.
OpenSOC-AI represents a practical, reproducible step in that direction.

\section*{Acknowledgment}

The author thanks the University of Alabama at Birmingham Department of
Computer and Information Sciences for academic support throughout this
project.


\appendix
\section{Training Configuration}

\begin{table}[htbp]
\caption{Full Training Hyperparameter Configuration}
\label{tab:config}
\begin{center}
\begin{tabular}{ll}
\toprule
\textbf{Parameter} & \textbf{Value} \\
\midrule
Base Model              & TinyLlama/TinyLlama-1.1B-Chat-v1.0 \\
Quantization            & 4-bit NF4, double quant., fp16 compute \\
LoRA rank ($r$)         & 16 \\
LoRA alpha              & 32 \\
LoRA dropout            & 0.05 \\
Target modules          & q, k, v, o, gate, up, down proj \\
Trainable params        & 12,615,680 (1.13\% of 1,112,664,064) \\
Training examples       & 450 \\
Eval examples           & 50 \\
Epochs                  & 3 \\
Per-device batch size   & 4 \\
Gradient accumulation   & 4 steps (effective batch = 16) \\
Learning rate           & 2e-4, cosine schedule, 5\% warmup \\
Optimizer               & paged\_adamw\_8bit \\
Max sequence length     & 512 tokens \\
Final training loss     & 0.0886 \\
Hardware                & NVIDIA Tesla T4 16 GB (Google Colab) \\
Training duration       & 4 minutes 21 seconds \\
Transformers version    & 4.40.2 \\
PEFT version            & 0.10.0 \\
TRL version             & 0.8.6 \\
bitsandbytes version    & 0.43.3 \\
\bottomrule
\end{tabular}
\end{center}
\end{table}

\end{document}